\documentstyle[12pt]{article}

 \def\appendix{\par
 \setcounter{section}{0}
 \setcounter{subsection}{0}
 \def\thesection{\Alph{section}}
 \def\theequation{\thesection.\arabic{equation}}}
 \def\thebibliography#1{\subsection*{References}\list
 {[\arabic{enumi}]}{\settowidth\labelwidth{[#1]}
 \leftmargin\labelwidth
 \advance\leftmargin\labelsep
 \usecounter{enumi}}
 \def\newblock{\hskip .11em plus .33em minus .07em}
 \sloppy\clubpenalty4000\widowpenalty4000
 \sfcode`\.=1000\relax}

\def\a{\alpha}
\def\b{\beta}
\def\c{\chi}
\def\d{\delta}

\def\g{\gamma}
\def\h{\eta}
\def\j{\psi}
\def\k{\kappa}
\def\l{\lambda}
\def\m{\mu}

\def\q{\theta}
\def\r{\rho}
\def\s{\sigma}
\def\t{\tau}
\def\ups{\upsilon}
\def\x{\xi}
\def\z{\zeta}

\def\F{\Phi}
\def\G{\Gamma}
\def\J{\Psi}
\def\Ld{\Lambda}

\def\X{\Xi}

\def\cc{{\cal C}}
\def\cd{{\cal D}}

\def\ck{{\cal K}}

\def\cy{{\cal Y}}

\def\inbar{\vrule height1.5ex width.4pt depth0pt}
\def\rlx{\relax\leavevmode}
\def\I{\leavevmode\hbox{\small1\kern-3.8pt\normalsize1}}
\def\openone{\leavevmode\hbox{\small1\kern-3.3pt\normalsize1}}
\def\Ione{\rlx{\rm 1\kern-2.7pt l}}
\font\cmss=cmss10
\font\cmsss=cmss10 at 7pt
\def\ZZ{\rlx\leavevmode
             \ifmmode\mathchoice
                    {\hbox{\cmss Z\kern-.4em Z}}
                    {\hbox{\cmss Z\kern-.4em Z}}
                    {\lower.9pt\hbox{\cmsss Z\kern-.36em Z}}
                    {\lower1.2pt\hbox{\cmsss Z\kern-.36em Z}}
               \else{\cmss Z\kern-.4em Z}\fi}
\def\Ik{\rlx{\rm I\kern-.18em k}}  % Yes, I know. This ain't capital.
\def\IC{\rlx\leavevmode
             \ifmmode\mathchoice
                    {\hbox{\kern.33em\inbar\kern-.3em{\rm C}}}
                    {\hbox{\kern.33em\inbar\kern-.3em{\rm C}}}
                    {\hbox{\kern.28em\sinbar\kern-.25em{\rm C}}}
                    {\hbox{\kern.25em\ssinbar\kern-.22em{\rm C}}}
             \else{\hbox{\kern.3em\inbar\kern-.3em{\rm C}}}\fi}
\def\IP{\rlx{\rm I\kern-.18em P}}
\def\IR{\rlx{\rm I\kern-.18em R}}
\def\IN{\rlx{\rm I\kern-.20em N}}

\newcommand{\ie}{{{\em i.e.}\ }}

\def\llsymbol#1{\@llsymbol{\@nameuse{c@#1}}}
\def\@llsymbol#1{\ifcase#1\or {}\or {'}\or {''}\or {'''}\or
   {''''}\or {'''''}\or  \else\@ctrerr\fi\relax}

\newcounter{contador}

\def\acknowledgement{\if@twocolumn
\section*{Acknowledgements}
\else \normalsize
\begin{center}
{\bf Acknowledgements\vspace{-.5em}\vspace{0pt}}
\end{center}
\quotation
\fi}
\def\endacknowledgement{\if@twocolumn\else\endquotation\fi}

\newcommand{\ol}\overline
\newcommand{\ti}\tilde
\newcommand{\wt}\widetilde
\newcommand{\wh}\widehat
\newcommand{\bv}\breve
\newcommand{\dg}\dagger

\newcommand{\C}{^{\mbox{\scriptsize c}}}

\newcommand{\QED}{QED$_{\mbox{\scriptsize 2+2}}$}

\newcommand{\aand}{\;\;\;\mbox{and}\;\;\;}

\newcommand{\be}{\begin{equation}}
\newcommand{\ee}{\end{equation}}
\newcommand{\bl}{\begin{eqnarray}&}
\newcommand{\el}{&\end{eqnarray}}
\newcommand{\bq}{\begin{eqnarray}}
\newcommand{\eq}{\end{eqnarray}}
\newcommand{\0}{{\bf 0}}

\newcommand{\sz}{\sigma_z}

\newcommand{\ad}{{\dot\alpha}}
\newcommand{\bd}{{\dot\beta}}

\newcommand{\uptad}{\widetilde\theta^{\dot\alpha}}

\newcommand{\qwt}{\widetilde\theta}
\newcommand{\ov}{\overline}
\newcommand{\pa}{\partial}

\def\sl#1{\rlap{\hbox{$\mskip 1 mu /$}}#1}	% good slash for lower case
\begin{document}

\title{\bf Non-Linear Supersymmetric $\sigma $-Models and their Gauging
in the Atiyah-Ward Space-Time}
\author{{\it M. Carvalho}\thanks{Internet e-mail:
kitty@cbpfsu1.cat.cbpf.br} ,~{\it L.C.Q. Vilar}{\thanks{Internet
e-mail: lcqvilar@cbpfsu1.cat.cbpf.br}} \\
{\it and} \\
 {\it J.A. Helay\"el-Neto } \\
Centro Brasileiro de Pesquisas F\'\i sicas (CBPF) \\
Departamento de Teoria de Campos e Part\'\i culas (DCP)\\
Rua Dr. Xavier Sigaud, 150 - Urca \\
22290-180 - Rio de Janeiro - RJ - Brazil.}

\date{}

\maketitle

\begin{abstract}

We present a supersymmetric non-linear $\s$-model built up
in the $N=1$ superspace of Atiyah-Ward space-time.
A manifold of the K\"ahler type comes out that is restricted
by a particular decomposition of the K\"ahler potential. The
gauging of the $\s$-model isometries is also accomplished in
superspace.

\end{abstract}

\newpage

\section{Introduction}

 In the past few years, there has been a great deal of attention
drawn
to the formulation of globally and locally supersymmetric models
 in Atiyah-Ward
space-times. One expects that self-dual (super) Yang-Mills theories
in $D=(2+2)$ might act as a potential source of new examples of
integrable
models \cite{atiyah,gates1}. Besides, it is well known that
 Atiyah-Ward space-times are the critical target
manifolds for string models with 2 supersymmetries in the
 world-sheet \cite{vafa} and that they also provide actions
for N=1 and N=2 supersymmetric non Abelian Chern Simons theory
in $D=(2+1)$ from a suitable dimensional reduction
of a self dual super Yang-Mills theory \cite{nish}.

Supersymmetry in $D=(2+2)$ reveals a number of peculiarities as
 due to the
special properties of spinors in such a space: Majorana-Weyl spinors
 may be
defined \cite{gates2} and
contrary to the case of $D=(3+1)$, the chirality constraint in
superspace
is not
affected by complex conjugation of superfields.
 This statement
is crucial in the process of building up actions for the matter
sector:
propagation is achieved only if independent superfields with
 opposite
chiralities mix together \cite{oswaldo}.

This property of mixing different chirality sectors that are not
related
to one
another by means of a simple complex conjugation has a major
influence
on the
coupling to Yang-Mills superfields as well as on the formulation of
supersymmetric non-linear $\s$-models. These models, in $D=(3+1)$
dimensions, have
played an important role in the coupling of supersymmetric gauge
 theories
 to supergravity. This was due to the non linear nature of the
coupling
 in a supergravity model, that can be interpreted in terms of a
 supersymmetric non-linear $\s$-model \cite{bagger}.

In the present work, we aim at an analysis of the geometrical
properties
of manifolds that may underline the construction of supersymmetric
non-linear $\s$-models in $D=(2+2)$, as much as possible very close
to the study of the
strong connection that exists between complex manifolds and
supersymmetries defined on space-times with a single timelike
coordinate \cite{zumino,roczek,freedman,hitchin,jose}.
However, working in the Atiyah-Ward space-time brings new features
to those formulations. Especifically, in the $N=1$ formulation of
 the supersymmetric $\s$-model in terms of a K\"ahler manifold we
 will be
led to assume it as a 4n dimensional manifold, its K\"ahler potential
being constrained by a certain decomposition. This naturally restricts
 our manifold to a subclass of
the more general possible K\"ahler manifolds. Our work is organized
as follows: in Section 2, we discuss the superspace
formulation of the model and establish its connection to K\"ahler
 manifolds.
In Section 3, we contemplate the description of isometries and
geometrical
conditions are set that allows us to conclude whether or not there
 will be
obstructions to the gauging of the isometries. The latter is the
 subject of
Section 4, where we also perform the coupling of the $\s$-model
to the
Yang-Mills sector of $N=1$, $D=(2+2)$ supersymmetry. The procedure
adopted in Sections 2 and 3 follows strongly that of \cite{roczek}.
Finally, our Concluding
Remarks are cast in Section 5.
An Appendix follows, where we set up some useful remarks about Killing
structures in our K\"ahler space.

\section{The Model in Superspace}

In our construction, we will closely follow the method used by Zumino
{\cite{zumino}}
 for deriving a supersymmetric $\sigma-$model action in
$D=(3+1)$ dimensions.
Here, the scalar fields defining the $\sigma-$model are the lowest
components of a
 set of chiral and antichiral superfields,
$(\F^i,\X^i) (i=1...n)$, which in $D=(2+2)$ are conveniently written
as
(we adopt the notation and conventions of Ref. \cite{oswaldo})
\be
\F^{i}= A^{i} + i\q\j^{i} + i\q^{2} F^{i} +
i\tilde\q \tilde{\sl{\pa}}\q A^{i}
+\frac{1}{2}\q^{2}\tilde\q \tilde{\sl{\pa}}\j^{i}
-\frac{1}{4}\q^{2}\tilde\q^{2}\Box A^{i} \;\; ,
\ee

\be
\X^{i}= B^{i} + i\tilde\q\tilde\c^{i} + i\tilde\q^{2} G^{i}
 + i\q \sl{\pa}
\tilde\q B^{i}
 +\frac{1}{2}\tilde\q^{2}\q
\sl{\pa} \tilde \c^{i} -\frac{1}{4}\q^{2}\tilde\q^{2}\Box B^{i}\;\;\;,
\ee
where $A,\;B$ are complex scalars, $\j,\;\c$ are Weyl spinors
 and $F,\;G$
 are complex scalar auxiliary fields.
It should be noted that, contrary to the $D=(3+1)$ case, complex
conjugation does not change chirality, i.e.
\bq
&{\wt{D}_{\ad}}\F^{i}=0 \aand {\wt{D}_{\ad}}\F^{i*}=0 \;\; , \nonumber \\
&D_{\a}\X^{i}=0 \aand D_{\a}\X^{i*}=0 \;\; ,
\label{quiral}
\eq
with
\be
D_{\a}=\pa_{\a}-i\sl{\pa}_{\a \ad}\uptad\aand
{\wt{D}_{\ad}}=\wt{\pa}_{\ad}-i\sl{\wt{\pa}}_{\ad \a}\q^{\a}\;\;\;,
\label{derivatives}
\ee

\bq
&\{D_{\a},{\wt{D}_{\ad}}\}=-2i\;{\s}^{\m}_{\a
\ad}\;{\pa}_{\m}\;,\;\{D_{\a},{{D_{\b}}}\}=\{{\wt{D}_{\ad}},
{\wt{D}_{\bd}}\}=0
\;\;,
\nonumber \\
&[D_{\a},{\pa}_{\m}]=[{\wt{D}_{\ad}},{\pa}_{\m}]=0\;\; ,
\label{dalgebra}
\nonumber
\eq
$\F^{i*}(\X^{i*})$ being the complex conjugates of $\F^{i}(\X^{i})$.
Following Zumino, we take for the supersymmetric action
{\footnote{$\int{d^4xd^2{\q}d^2\qwt} \equiv \int{d^4x}D^{\a}
{\wt{D}^{\ad}}\wt{D}_{\ad}D_{\a}$}}
\be
S\;\;=\;\;\frac{1}{8}\int{d^4xd^2{\q}d^2\qwt}\;K(\F^{i},\X^{i};
\F^{i*},\X^{i*}) \;\;\; ,
\label{action}
\ee
where the potential $K$ is a real function. It is obvious from
 (\ref{action})
that
we need to take the manifold spanned by the scalar fields as a 4n
manifold. Terms involving only one chirality, e.g.,
functions of $\F^{i}$ and $\F^{i*}$ or $\X^{i}$ and
 $\X^{i*},$
 would not provide the kinetic term for the $\s-$model.
Then, from the component expansion of (\ref{action}),
 we get
\bq
S\;\;&=&\;\;2\;\int{d^{4}x}\; \biggl( \;\frac{\pa^2 K}
{\pa A^{i}\pa B^{j}}
\pa _{\m}A^{i}\pa^{\m}B^{j} +
\frac{\pa^2 K}{\pa A^{i}\pa {B^{*j}}} \pa_{\m}A^{i}\pa^{\m}{B^{*j}}
\nonumber \\
&&\;\;\;
+ \frac{\pa^2 K}{\pa {A^{*i}}\pa B^{j}}\pa _{\m}{A^{*i}}
\pa^{\m}B^{j} +\frac{\pa^2 K}{\pa {A^{*i}}\pa {B^{*j}}}
\pa _{\m}{A^{*i}}\pa^{\m}{B^{*j}} + interaction\;\; terms\; \biggr) \; .
\label{action1}
\eq
In the latter expression, we have written only the part
associated to the
kinetic term of the complete action, which gives us the metric of the
manifold as
\be
g_{{\cal I} {\cal J}} =
%\equiv \frac{\pa^2 K}{\pa X^{\cal I}
%\pa X^{\cal J}}=
\left(\begin{array}{cccc}
\0 &\frac{\pa^2 K}{\pa A^{i}\pa B^{j}} &\0
&\frac{\pa^2 K}{\pa A^{i}\pa B^{*j}} \\
\frac{\pa^2 K}{\pa B^{i}\pa {A^{j}}} &\0 &\frac{\pa^2 K}
{\pa B^{i}\pa A^{*j}}
&\0 \\
\0 &\frac{\pa^2 K}{\pa {A^{*i}}\pa B^{j}} &\0 &\frac{\pa^2 K}
{\pa {A^{*i}}\pa
B^{*j}} \\
\frac{\pa^2 K}{\pa B^{*i}\pa A^{j}} &\0 &\frac{\pa^2 K}
{\pa {B^{*i}}\pa A^{*j}}
&\0 \\
\end{array}\right)\;\;\;,
\label{met} \ee
where
\bq
%&\left\{\begin{array}{l}
%X_{\cal I}=(X_{i},X_{n+i},X_{2n+i},X_{3n+i}) =
%%(A_{i},B_{i},{A_{i}}^*,{B_{i}^*})
%\nonumber \\
{\cal I},{\cal J}=1,...4n\;\;and\;\; i,j=1,...n\;\;.
%\end{array}\right.
\nonumber
\eq
Equation (\ref{met}) shows that in a four-dimensional space-time
 with signature 2+2,
 it is not necessary that a supersymmetric $\s-$model
be associated with a K\"ahler manifold, contrary to what happens in
$D=(3+1)$. In fact, a condition for having a K\"ahler metric is that
$g_{{\cal I} {\cal J}}$ should be hybrid
{\cite{Yano}} and here this can only be achieved if $K$ may be
decomposed as
below:
\be
K(\F^{i},\X^{i};\F^{i*},\X^{i*}) \;\;\; = \;\;\;
 H(\F^{i},\X^{i*}) \;\; + \;\;
H^*(\F^{i*},\X^{i})\;\;\;.
\label{deco} \ee
Consequently, if this is the case, the metric turns out to be
\be
g_{{\cal I} {\cal J}}=
\left(\begin{array}{cc}
\0
&g_{I {\ov J}} \\
g_{{\ov I} J} &\0 \\
\end{array}\right)=
\left(\begin{array}{cccc}
\0 &\0 &\0
&\frac{\pa^2 H}{\pa A^{i}\pa {B^{j}}^*} \\
\0 &\0 &\frac{\pa^2 H^*}{\pa B^{i}\pa {A^{j}}^*} &\0 \\
\0 &\frac{\pa^2 H^*}{\pa {A^{i}}^*\pa B^{j}} &\0 &\0 \\
\frac{\pa^2 H}{\pa {B^{i}}^*\pa A^{j}} &\0  &\0 &\0 \\
\end{array}\right)\;\;\;.
 \label{metric}
\ee
In the above expression for $g_{{\cal I} {\cal J}}$, we suceeded in
explicitly
writing down the off-diagonal structure that characterizes the metric
for
K\"ahler manifolds \cite{Yano}. But as the holomorphic structure has
 been
partitioned in two disjoint pieces by (\ref{deco}), we can conclude
that
the manifold we arrived at is in fact more constrained than a general
K\"ahler manifold. This will become clearer in the next section
when we
will discuss the isometries of this manifold.

It is also interesting to notice that the four-block K\"ahlerian
structure in
(\ref{metric}) resembles that of a Hyper-K\"ahler space, although here we
do not have the other complex structures (or equivalently, the second
supersymmetry) which characterizes this latter space.  The analysis of
such $N=2$ models will be presented elsewhere \cite{nosso}.

With this choice for the potential $K$ and using the equations of
 motion
to eliminate the auxiliary fields, we get from
(\ref{action}) the full action as
\bq
S &=& \int{d^{4}x}\;
\biggl( \;\;2\;h_{i \ov{\hat j}}\pa_{\m}A^{i}\pa^{\m}{B^{*j}}
+2\;h^{*}_{\ov{i} \hat j}\pa_{\m}{A^{*i}}\pa^{\m}B^{j}
-\frac{1}{2}i\;h_{i \ov{\hat j}}\ti\c^{cj}\ti\s^{\m}\cd_{\m}\j^{i}
 \nonumber \\
&& \;\;\; -\frac{1}{2}i\;h_{i \ov{\hat j}}\j^{i}\s^{\m}\cd_{\m}
\ti\c^{cj}
 -\frac{1}{2}i\;h^{*}_{\ov{i} \hat j}\ti\c^{j}\ti\s^{\m}
\cd_{\m}\j^{ci}
-\frac{1}{2}i\;h^{*}_{\ov{i} \hat j}\j^{ci}\s^{\m}\cd_{\m}\ti\c^{j}
\nonumber \\
&&\;\;\;-\frac{1}{8}(\;h^{k \ov{\hat l}}\pa_{\ov {\hat i}}h_{k \ov {\hat
j}}\pa_{m}h_{n \ov {\hat l}}-\pa_{m}\pa_{\ov {\hat i}}h_{n\ov {\hat
j}})\ti\c^{ci}\ti\c^{cj}\j^{m}\j^{n} \nonumber \\
&&\;\;\; -\frac{1}{8}(\;h^{* \ov{k} \hat{l}}\pa_{\hat i}h^{*}_{\ov k
{\hat j}}\pa_{\ov m}h^{*}_{\ov n {\hat l}}-\pa_{\ov m}\pa_{\hat i}
h^{*}_{\ov n {\hat  j}})\ti\c^{i}\ti\c^{j}\j^{cm}\j^{cn} \biggr)\;\;\;,
\label{action2} \eq
where we have denoted
\bq
\left\{\begin{array}{l}
\hat i = i+n ,\;\;\ov i= i+2n ,\;\;and\;\; \ov{\hat i} = i+3n \nonumber \\
X_{\hat i}=B_{i},\;\;X_{\ov i}={A_{i}}^*,\;\; X_{\ov{\hat i}}= {B_{i}}^*\;\;.
\end{array}\right.
\eq
The components of the metric were written as
\be
h_{i \ov{\hat j}}=\frac{\pa^2 H}{\pa A^{i}\pa {B^{*j}}}\;\; ,\;\;\;\;
h^{*}_{\hat i \ov{j}}=\frac{\pa^2 H^{*}}{\pa B^{i}\pa {A^{*j}}}\;\;
 ,\;\;\;\;
h^{*}_{\ov i {\hat j}}=\frac{\pa^2 H^{*}}{\pa A^{*i}\pa {B^{j}}}\;\;
,\;\;\;\;
h_{\ov{\hat i} j}=\frac{\pa^2 H}{\pa B^{*i}\pa {A^{j}}}\;\;,
\nonumber \ee
and the covariant derivatives for the fermions are directly read off:
\bq
\left \{\begin{array}{l}
\cd _{\m}\j^{i}=\pa_{\m}\j^{i}+
h^{i\ov {\hat  l}}\pa_{k}h_{j\ov {\hat  l}}\j^{k}\pa_{\m}A^{j}\;\; , \nonumber
\\
\cd _{\m}\ti\c^{ci}=\pa_{\m}\ti\c^{ci}+
h^{l\ov {\hat  i}}\pa_{\ov {\hat k}}h_{l\ov {\hat
j}}\ti\c^{ck}\pa_{\m}B^{*j}\;\; ,
\nonumber \\
\cd _{\m}\j^{ci}=\pa_{\m}\j^{ci}+h^{*{\ov i}\hat l}\pa_{\ov k}
h^{*}_{{\ov j}\hat  l}\j^{ck}\pa_{\m}A^{*j}\;\; , \nonumber \\
\cd _{\m}\ti\c^{i}=\pa_{\m}\ti\c^{i}+
h^{*{\ov l}{\hat  i}}\pa_{\hat k}h^{*}_{{\ov l} \hat
j}\ti\c^{k}\pa_{\m}B^{j} \;\; .
\nonumber
\end{array} \right .
\eq
In the above expressions $\j\C$$\equiv$$i\sz\j^*$ and
$\wt\c\C$$\equiv$$i\sz\wt\c^*$ \cite{oswaldo}. Using them we get,

\bq
-\frac{1}{2}i\;h_{i \ov{\hat j}}\ti\c^{cj}\ti\s^{\m}\cd_{\m}\j^{i}
-\frac{1}{2}i\;h^{*}_{\ov{i} \hat j}\ti\c^{j}\ti\s^{\m}\cd_{\m}\j^{ci}
 &=&
2Re \biggl\{-\frac{1}{2}i\;h_{i \ov{\hat j}}\ti\c^{cj}\ti\s^{\m}\cd_{\m}\j^{i}
\biggr \}
\nonumber \\
\biggl( \ti\c^{ci}\ti\c^{cj}\j^{m}\j^{n}\biggr)^*
  &=&
\c^{i}\c^{j}\j^{cm}\j^{cn} \nonumber
\eq
from which we can easily conclude for the reality of the action.

We can get a very simplified expression if we introduce
{\footnote{We will also use $Z^{I}$ as denoting
$Z^{I}=(\F^{i},\X^{i}).$}
\bl
\J^{I}_{A}=
\left(\begin{array}{c}
 \j^{i}_{\a} \\
\ti \c^{i}_{\dot \a}
\end{array}\right), \;\;\;
%\aand
\J^{\ov I}_{A}=
\left(\begin{array}{c}
 \j^{ci}_{\a} \\
\ti \c^{ci}_{\dot \a}
\end{array}\right),\;\;\;
Z^{I}=
\left(\begin{array}{c}
 A^{i} \\
 B^{i}
\end{array}\right), \;\;\;
Z^{\ov I}=
\left(\begin{array}{c}
 A^{*i} \\
 B^{*i}
\end{array}\right) \;\; ,
\label{compact}
\el
and the matrix
%\be
%\s_{AB}^{\m}=
%\left(\begin{array}{cc}
%\ti\s^{\m}_{\dot \a \b} &\0 \\
%\0 &\s^{\m}_{\a \dot \b} \\
%\end{array}\right) \;\; (A=\{\a,\dot\a\},\;\; B=\{\b,\dot\b\})
%\ee

\be
\g_{AB}^{\m}=
\left(\begin{array}{cc}
\0 &\s^{\m}_{\a \dot \b} \\
\ti\s^{\m}_{\dot \a \b} &\0 \\
\end{array}\right) \;\; (A=\{\a,\dot\a\},\;\; B=\{\b,\dot\b\})
\ee
Then the action (\ref{action2}) becomes
\bq
S &=& \int{d^{4}x}\;
\biggl( 2 g_{I \ov J}\;\pa_{\m}Z^{I}\pa^{\m}Z^{\ov J}
-\frac{i}{2} g_{{\ov I} J}\;\J^{\ov I}\g^{\m}\cd_{\m}\J^{J}
-\frac{i}{2} g_{I \ov J}\;\J^{I}\g^{\m}\cd_{\m}\J^{\ov J}
\nonumber \\
 && \;\;\;\; +\frac{1}{8}R_{I\ov M J\ov N}\;\J^{\ov M}\J^{\ov N}\J^{I}\J^{J}
\biggr)\;\;\;,
\eq

%\J^{\hat j}\tilde{\s^{\m}}\cd_{\m}\J^{i}
where
\bq
\cd_{\m} \J^{IA}&=&\pa_{\m}\J^{IA} + g^{I \ov L}\;\pa_{K} \;
 g_{J \ov L}\;\J^{KA}\pa_{\m}Z^{J} \;\; , \nonumber \\
 R_{I\ov M J\ov N}&=& \pa_{I}\pa_{\ov M}\;g_{J \ov N}-
g^{\ov K L}\;\pa_{I}\;g_{\ov K J} \; \pa_{\ov M}\;g_{\ov N L}
\label{riemann} \;\;,
\eq
this expression being similar in form with the action appearing in
\cite{zumino}.

It should be noted that in $D=(2+2)$, we can also formulate a non-linear
supersymmetric $\s-$model using chiral and anti-chiral real superfields
$(\F^{i}=\F^{i*},\;\; \X^{i}=\X^{i*})$. Here, we take our  potential
$K$
as a function of $(\F^{i},\X^{i})$ and the action as in the usual form

\be
S\;\;=\;\;\frac{1}{8}\int{d^4xd^2{\q}d^2\qwt}\;K(\F^{i},\X^{i})\;\;.
\ee
We obtain
\be
S\;\;=\;\;\int{d^{4}x}\;
\biggl( \;\;2 g_{i{\hat j}}\pa_{\m}Z^{i}\pa^{\m}Z^{\hat j}
-\frac{i}{2}g_{{\hat i}j}\J^{\hat i}\tilde{\s^{\m}}\cd_{\m}\J^{j}
-\frac{i}{2}g_{i{\hat j}}\J^{i}\s^{\m}\cd_{\m}\J^{\hat j}
+\frac{1}{8}R_{i{\hat m}j{\hat n}}\J^{\hat m}\J^{\hat
n}\J^{i}\J^{j}\biggr)\;\;,
\ee
where we have used a notation similar to (\ref{compact}), but with the hated
components denoting the chiral conjugates, \ie $(Z^{i},Z^{\hat i}) \equiv
(A^{i},B^{i})$ and $(\J^{i},\J^{\hat i}) \equiv (\j^{i},\ti\c^{i})$. Naturally,
the
metric is
\bq
g_{i{\hat j}}=
\frac{\pa^2 K}{\pa A^{i}\pa B^{j}} \;\; ,
\nonumber
\eq
and the covariant derivatives and Riemann curvature are  totally
analogous to (\ref{riemann}).

Obviously, this space is not K\"ahlerian, as it is not a complex space.
 But it is curious to notice that it possesses some properties of a
K\"ahler space if we just replace the notion of complex conjugation
by that of chiral conjugation, that would take $Z^i$ into $Z^{\hat i}$
and vice-versa. This space is a feature of the $2+2$ signature. In
$D=(3+1)$, a $N=1$ supersymmetric $\s$-model requires a complex K\"ahler
manifold as target \cite{zumino,freedman}. We are seeing then another
example, together with that appearing in (\ref{met}), of a non-K\"ahler
manifold associated to $N=1$ supersymmetric $\s$-models in $D=(2+2)$,
although having the nice feature of being included in the class of
theories generated by a potential $K$.

%with,
%\bq
%\cd_{\m}\J^{i}=\pa_{\m}\J^{i} + g^{i\hat l}\pa_{j}g_{k\hat
%%%l}\J^{k}\pa_{\m}A^{j}\nonumber \\
%\cd_{\m}\J^{\hat i}=\pa_{\m}\J^{\hat i} + g^{{\hat i} l}\pa_{\hat %j}g_{\hat k
%l}\J^{\hat k}\pa_{\m}B^{j}\nonumber \\
%R_{\hat i m\hat j n} = \pa_{\hat i}\pa_{m}g_{\hat j n} - g^{k\hat %l}\pa_{\hat
%%r}g_{k\hat j}\pa_{m}g_{n\hat l}
%\eq

\section{Isometries}

In the previous section, we have imposed the decomposition (\ref{deco})
in order to render manifest the K\"ahlerian structure of the target space. From
(\ref{metric}), we observe that the transformations
for the potential $K$, allowed by the condition of metric invariance, are
of the form

\be
K \;\;\; \longrightarrow \;\;\; K^{'} \;\;=\;\; K\;+\; F(Z_I)\; +\,
G(Z_{\ov I}) \;\;.
\label{t1} \ee
These are the holomorphic transformations of a general K\"ahler manifold.
 Nevertheless, the $D=(2+2)$ spacetime structure forbids such a
transformation, since terms out of the blocks $g_{I {\ov J}}$ would be
generated. This happens because the invariance of the action (\ref{action})
 is ensured by chiral transformations

\be
K \;\;\; \longrightarrow \;\;\; K^{'} \;\;=\;\; K\;+\; F(\F, \F^* )\;
+\, G(\X, \X^* ) \;\;.
\label{t2} \ee
%We remind that in D=(2+2), $\F$ and $\F^{*}$ are both chiral, whereas $\X$ and
%$\X^{*}$ are anti-chiral, see eq.(\ref{quiral}).
The way to make (\ref{t1})
and (\ref{t2}) compatible is to admit that the most
general transformation of the potential $K$ is

\be
K \;\;\; \longrightarrow \;\;\; K^{'} \;\;=\;\; K\;+\; \h (\F)
\; + \; \h^{*}(\F^* )\;
+\, \q (\X) \; + \; \q^{*} (\X^* ) \;\;.
\label{trans1} \ee
This has an immediate consequence on the possible coordinate
transformations allowed for the manifold. The holomorphic transformations of
a general K\"ahler manifold are decomposed into a more restricted subgroup
in which coordinates associated to different chiralities do not mix
\footnote{ From now on, the term holomorphic will mean not only a splitting
in terms of fields and their conjugated, but also a splitting in different
chiralities.}
%from now on the term holomorphic will refer not only to quantities that splits
%%%in terms of complex conjugation but also in terms of chirality.
\be
A^{i}\; \longrightarrow \; A^{'i} = f(A^{i}) \;\;\;\; ,  \;\;\;\;
B^{i}\; \longrightarrow \; B^{'i} = f(B^{i}) \;\;\;\;\; and \;\; c.c.
\;\; .
\ee
If we permitted that a coordinate $ A^{i}$ could have been taken into a
$ B^{i}$, terms out of the anti-diagonal in the metric of (\ref{metric})
would have been
generated. In this way, we see that we are dealing with a subset of manifolds
among those that have the most general K\"ahler form. Also, from these facts,
we can conclude that the Killing vectors will
be parametrized in terms of different chiral components:
\be
{\cal K}^{I}_{a} \;\; = \;\; \left(\begin{array}{c}
{\k}^i_a (A) \nonumber \\
{\t}^{i}_a (B) \end{array}\right) \;\;\; , \;\;\;
{\cal K}^{\ov I}_{b} \;\; = \;\; \left(\begin{array}{c}
{\k}^{*i}_b (A^*) \nonumber \\
{\t}^{*i}_b (B^*) \end{array}\right) \;\; .
\label{killing1} \ee
The possibility of working with the above Killing vectors is due to the fact
that the metric does not contain the components $g_{i{\ov j}},
\;g_{{\hat i}\ov{\hat j}},\;g_{{\ov i}j},\;g_{{\ov{\hat i}}\hat j}\;$
 (see Appendix).
Under a global isometry, the coordinates of the K\"ahler manifold
will transform as
\be
Z^{'I} \; = \; \exp{(L_{\l \cdot \cal K})} Z^{I} \rightarrow
\left\{\begin{array}{l}
A^{'i}\; =\; \exp{(L_{\l \cdot \k})} A^{i}
\nonumber \\
B^{'i} \; = \; \exp{(L_{\l \cdot \t})} B^{i}
\end{array} \right.
\;\; and \;\; c.c.
\;\; ,
\label{isoco}
\ee
where $L_{X}$ is the Lie derivative along X and $\l$ is a global parameter.
 The Killing vectors
generate the algebra  of the isometry group of the K\"ahler manifold, i.e.
$[{\cal K}_{a},{\cal K}_{b}]= f_{ab}^c \;{\cal K}_c.$
The isometries induce transformations in
the potential $K$, which are described in their general form by
\be
\d K \;\;
= \;\; {\l}^a  \biggl( \frac{\pa K}{\pa Z_I} {\cal K}^I_a \;\; + \;\;
\frac{\pa K}{\pa Z_{\ov I}} {\cal K}^{\ov I}_a \biggr) \;
= \;
{\l}^a \frac{\pa H}
{\pa A^i} {\k}^i_a +
{\l}^a \frac{\pa H^*}
{\pa B^i} {\t}^{i}_a +
{\l}^a \frac{\pa H^*}
{\pa A^{*i}} {\k}^{*i}_a +
{\l}^a \frac{\pa H}
{\pa B^{*i}} {\t}^{*i}_a  \;\; .
\label{trans2} \ee
Comparing eq.(\ref{trans2}) with eq.(\ref{trans1}), which
is also an invariance of the metric, we can write
\bq
\h_a (A) \; & = &\; \frac{\pa H(A,B^*)}
{\pa A^i} \; {\k}^i_a (A) \; + \; Y_a (A,B^*) \;\; , \nonumber \\
\q_a (A) \; & = &\; \frac{\pa {H^*}(B,A^*)}
{\pa B^i} \; {\t}^i_a (B) \; - \; Y^*_a (B,A^*) \;\; , \nonumber \\
\h^*_a (A^*) \; & = & \; \frac{\pa H^* (B,A^*)}
{\pa A^{*i}} \; \k^{*i}_a (A^*) \; + \; Y^*_a (B,A^*) \;\; , \nonumber \\
\q^*_a (A^*) \; & = & \; \frac{\pa H(A,B^*)}
{\pa B^{*i}} \; \t^{*i}_a (B^*) \; - \; Y_a (A,B^*) \;\; .
\label{relisom} \eq
The introduction of the complex functions $Y_a$ is necessary,
so that no further restriction is imposed on the potentials $H's$.
The $Y_a's$ are naturally related to the structure of Killing vectors
in K\"ahler space. To show this, we may start by the derivation
of the first equation in (\ref{relisom}) with respect to $B^*$:
\be
 \frac{ {\pa}^2 H}
{\pa A^i \pa B^{*j}} \; {\k}^i_a (A) \; = \;
 - \; \frac{\pa {Y_a}}{\pa B^{*j} } \;\; ,
\ee
or deriving $\q_{a}$ with respect to $A^*$:
\be
 \frac{ {\pa}^2 H^*}{\pa B^i \pa A^{*j}}  \; {\t}^i_a (B) \; = \;
 \;\; \frac{\pa {Y^*_a}}{\pa A^{*j} } \;\; .
\ee
These equations and their conjugates can be written in a
compact form in terms of a real potential ${\cy}_a = i Y^*_a (B,A^*) -
i Y_a (A,B^*)$
\be
g_{I \ov J} \; {\cal K}^I_a \; = \; - i \; \frac{\pa {{\cy}_a}}{\pa Z^{\ov
J}}
 \;\;\;\;\; and \;\; c.c.
\;\; .
\label{kx}
\ee
This equation is just the restriction imposed by the Killing equation with
 mixed indices,
\be
\nabla_I \; {\ck}_{\ov J} \; + \; \nabla_{\ov J} \; {\ck}_I \;\; = \;\; 0\;\;,
\ee
on the form of the Killing vectors, which become described by the
potential $\cy_a$.

The determination of this potential is crucial for the process
of gauging, as we shall see in what follows. In order to accomplish
this goal, we will use the method stablished in \cite{roczek}.
 Contracting eq.(\ref{kx}) with
 ${\cal K}^{\ov J}_b$ and its
conjugate with  ${\cal K}^{I}_b$, and then
comparing them both, we get the identity
\be
{\cal K}^{I}_b \; \frac{\pa \cy_a}{\pa Z^I}
\; + \; {\cal K}^{\ov I}_a  \; \frac{\pa \cy_b}{\pa Z^{\ov I}}
 \;\; = \;\; 0 \;\; .
\label{iden}\ee
Now, under an isometry transformation, $\cy_a$ transforms as
\be
\d \cy_a \;\;
= \;\; {\l}^b \; \biggl( \frac{\pa \cy_a}{\pa Z^I} \; {\cal K}^I_b \;\; + \;\;
\frac{\pa \cy_a}{\pa Z^{\ov I}} \; {\cal K}^{\ov I}_b \biggr) \;\; ,
\ee
which, by virtue of (\ref{iden}) may be written as
\be
\d \cy_a \;\;
= \;\; \frac{{\l}^b}{2} \; \biggl( \frac{\pa \cy_{[a}}{\pa Z^I} \;
{\cal K}^I_{b]} \;\; + \;\;
\frac{\pa \cy_{[a}}{\pa Z^{\ov I}}  \; {\cal K}^{\ov I}_{b]} \biggr) \;\; .
\label{deltay}\ee
Through eq.(\ref{trans2} - \ref{kx}) we get the fundamental relation
\be
 {\cal K}^I_{[a} \; \frac{\pa \cy_{b]}}{\pa Z^I} \;\; + \;\;
 {\cal K}^{\ov I}_{[a} \;  \frac{\pa \cy_{b]}}{\pa Z^{\ov I}}
\;\; = \;\; f_{ab}^c \; ( \x_c \; + \; \x^*_c ) \;\; ,
\ee
where $\x_a = \h_a + \q_a $ and $f_{ab}^c$ are the structure constants of
the isometry group. In components, this last equation means
 \bq
 {\k}^i_{[a} \;\frac{\pa \h_{b]}}{\pa A^i} \; & = & \;
 f_{ab}^c \; \h_c \; + \; c_{ab}  \nonumber \\
 {\t}^i_{[a} \; \frac{\pa \q_{b]}}{\pa B^i} \; & = & \;
 f_{ab}^c \; \q_c \; - \; c^*_{ab} \;\;\;\;\;\;\;\;\; and \;\;c.c.
\;\; ,
\label{comp}
\eq
and  $c^*_{ab} \; = \; - \; c^*_{ba} $ is a complex constant.
Finally, from (\ref{deltay} - \ref{comp}) we get
\be
\d Y_a \;\;
= \;\; {\l}^b \; \biggl( \frac{\pa Y_a}{\pa A^i} \; {\k}^i_{b}  \;\; + \;\;
\frac{\pa Y_a}{\pa B^{*i}} \; {\t}^{*i}_b \biggr) \;\; = \;\;
- \; {\l}^b \;  f_{ab}^c \; Y_c \; - \; {\l}^b \; c_{ab} \;\;\;\; and
\; c.c.
\;\; .
\label{delty}
\ee
At this point we see that, in order to make explicit the potentials
$Y_a$ as  functions of the Killing vectors, we have to restrict
the isometry groups to be semi-simple. This becomes clearer
if we combine (\ref{kx}) in (\ref{delty}):
\be
Y_a \; = \; 2 \;  f_{ab}^c \; \k^i_d \; \t^{*j}_c  \frac{ {\pa}^2 H}
{\pa A^i \pa B^{*j}}
\; g^{bd} \; + \; f_{ab}^c \; c_{dc} \; g^{bd} \;\;\;\; and \; c.c.
\;\; .
\label{y}
\ee
To define $Y_a$ we needed to introduce the inverse Killing metric, and this
means that Abelian factors would spoil the definition of $Y_{a}$, so that only
      semi-simple groups are allowed \cite{gilmore}. The constants $c_{ab}$
express an arbitrariness in the definition of $Y_a$, as they can be reabsorbed
by the shift $Y_a \; \rightarrow \; Y_a^{'} \;=\; Y_a\; - \;  f_{ab}^c c_{dc}
g^{bd}$, whenever $g^{ab}$ is defined. This property will be of fundamental
importance in the procedure of gauging the model.

In the particular case of a non-semi-simple group, $G$, of isometries,
for which  $f_{ab}^c$ is non-vanishing only when all its indices are
associated
to generators in the semi-simple factor $S$, \ie $G$ has the form
\be
G \;\; = \;\; S \;\; \otimes \;\; A_{N} \nonumber \;\; ,
\ee
where $A_{N}$ represents the direct product of $N$ Abelian factors,
and if all the constants $c_{ab}$ (determined by (\ref{comp})) with indices
associated to the latter vanish, then from (\ref{delty}) we can
conclude that
the potential $Y_{a}$ will be
always determined up to $N$ arbitrary complex constants associated to each
 Abelian factor.

In the general case of a non-semi-simple group, with Abelian factors
generating  non-zero constants $c_{ab}$, eq.(\ref{delty}) may not
admit any solution and this will be an obstruction to the gauging as we
shall see in the following.

\section{The Gauging}

The isometry transformations of the coordinates on a K\"ahler manifold
are given in eq.(\ref{isoco}). Now we can make this symmetry local by
taking the constant parameter $\l$ as superfields of definite chirality.
Those transformations are then written in superfields as
\bq
\F \longrightarrow \;\;\; \F^{'} \;\; & = & \;\; \exp{(L_{\Ld \cdot \k})} \F
\nonumber \\
\X \longrightarrow \;\;\; \X^{'} \;\; & = & \;\; \exp{(L_{\G \cdot \t})} \X
\;\; \;\;
and \;\; c.c. \;\; ,
\eq
The superfields $\Ld$ and $\G$ are chiral and anti-chiral respectively.
But as we have already seen, in $D=(2+2)$ this does not make any restriction
on their reality. In $D=(3+1)$ they would be necessarily complex conjugates of
each other.

%%%%%%%%%%%%%%%%%%%%%%%%%%%%%%%%%%%%%%%%%%%%%%%%%%%%%%%%%%
%So we have three possibilities
%\bq
%& i) & \;\;\;\; \Ld \; = \; \Ld^* \;\;\; , \;\;\; \G \; = \; \G^* \nonumber \\
%& ii) & \;\;\;\; \Ld \; \not=  \; \Ld^* \;\;\; , \;\;\; \G \; = \; \G^*
%%\nonumber %\\
%& iii) & \;\;\;\; \Ld \; \not=  \; \Ld^* \;\;\; , \;\;\; \G \; \not= \; \G^*
%\eq
%%%%%%%%%%%%%%%%%%%%%%%%%%%%%%%%%%%%%%%%%%%%%%%%%%%%%%%%%

Let us then take $\Ld \; = \; \Ld^*, \;\; \G \; = \; \G^*. $
Here, the local infinitesimal isometries read as
\bq \d\F^{i}  & = &  \Ld^{a}k_{a}^{i} \nonumber \\
 \d\X^{i} & = & \G^{a}\t_{a}^{i}  \;\; \;\;
and \;\; c.c. \;\; ,
\eq
and the K\"ahler potential transforms like

\be
\d K \;\;
= \;\;
{\Ld}^a \biggl(\frac{\pa H}
{\pa A^i} {\k}^i_a +
\frac{\pa H^*}
{\pa A^{*i}} {\k}^{*i}_a \biggr) +
\G^{a}\biggl( \frac{\pa H^*}
{\pa B^{i}} {\t}^{i}_a +
\frac{\pa H}
{\pa B^{*i}} {\t}^{*i}_a \biggr) \;\; .
\label{gauge1} \ee
In order to have a transformation which could be compared with
 (\ref{trans1}), all superfields should transform with the same parameter.
 This can be obtained if we introduce a real vector superfield $V$,
which in $D=(2+2)$ assumes the form,

\bq
V(x,\q,\qwt)\!\!\!&=&\!\!\!C(x)+i\q\z(x)+i\qwt\wt\h(x)+\frac12i\q^2M(x)+
\frac12i\qwt^2N(x)+\nonumber\\ &&+\frac12i\q\s^\m\qwt
A_\m(x)-\frac12\qwt^2\q\l(x)-\frac12\q^2\qwt\wt\r(x)-
\frac14\q^2\qwt^2 D(x)\;\;, \label{supervector}
\eq
where $C$, $M$, $N$ and $D$ are real scalars, $\z$, $\wt\h$, $\l$ and
$\wt\r$ are majorana-Weyl spinors and $A_\m$ is a vector field.
Now we replace the superfields $\X^{i}$ \cite{wessgris} by
\be
\ti\X^{i}\;\equiv \;\exp{(L_{V \cdot \t})}\X^{i} \;\;\; and \;\;\; c.c.
\;\; ,
\ee
 so that
 $\ti\X^{i}$  can transform as
\be
\ti\X ^{'i}\;=\;\exp{(L_{\Ld \cdot \t})}\ti\X^{i} \;\; .
\ee
This is only possible if the vector superfield
transforms as
\be
\exp{(L_{V^{'} \cdot \t})}\;=\;\exp{(L_{\Ld \cdot \t})}
\exp{(L_{V \cdot \t})}\exp{(-L_{\G \cdot \t})} \;\; .
\label{V}
\ee
Since the parameters $\Ld$ and $\G$ are real we have from (\ref{V}) that $V$
 really transforms as a real vector superfield.
The infinitesimal isometries have the form
\bq
\d\F^{i}  & = &  \Ld^{a}k_{a}^{i} \nonumber \\
 \d\ti\X^{i} & = & \Ld^{a}\t_{a}^{i}
 \;\;\;\;\;\;\;\;and \;\; c.c.
\;\; ,
\eq
and the transformation (\ref{gauge1}) assumes a form comparable to
 (\ref{trans1}),
with the replacements $ \{\X,\;\X^{*}\} \longrightarrow
\{\ti\X,\;\ti\X^{*}\}$
{}.
But now, since the parameter $\Ld$ is a chiral superfield, we do
not have the
action invariant under local isometries, for
\be
S \;\;\; \longrightarrow \;\;\;
S^{'}\;\;=\;\;\frac{1}{8}\int{d^4xd^2{\q}d^2\qwt}
\;\; \Ld^{a}\;\biggl(\q_{a}(\ti \X) \;
+ \; \q^{*}_{a}(\ti \X^{*})\biggr) \;\;
\neq  \;\; 0 \;\; .
\ee
However, the invariance of the action can be recovered if we introduce an
antichiral
 superfield and its complex conjugate, $\ups$ and $\ups^*$, such that they
transform like
\bq
&\d\ups\;\;=\;\;\l^{a}\q_{a}  \;\;(\X) , \nonumber \\
&\d \ups^*\;\;=\;\;\l^{a}\q^{*}_{a}(\X^*)\;\; .
\label{isoco1}
\eq
Then, we take our action as
\be
S_{\ups}\;\;=\;\;\frac{1}{8}\int {d^4xd^2{\q}d^2\qwt}\;
\biggl(H(\F,\X^{*}) \; + \;
H^{*}(\F^{*},\X) \;-\;\ups\;-\;
\ups^{*}\biggr)\;\;.
\label{Sv}
\ee
This action is globally invariant  under the infinitesimal form of
 the transformations (\ref{isoco}) and  (\ref{isoco1}), and since
$\ups\;$ and $\;\ups^{*}$
are anti-chiral, we also have $S_{\ups}\;=\;S$. Those superfields
should be thought of as extra coordinates extending
 our manifold \cite{roczek}. In
this way, we write two new Killing vectors
  \be
\t_{a}(\X) \;\longrightarrow \;\t_{a}^{'}(\X)\;=\;
 \t_{a}^{i}(\X) \frac{\pa}{\pa \X^i}\;+\;\q_{a}(\X)\frac{\pa}{\pa \ups}
\;\;\; and \;\;\;c.c. \;\; ,
\ee
and the new K\"ahler potential $K^{'}=K-\ups-\ups^{*}$ is invariant under
their action.
Finally, the gauging of the isometry is simply performed by replacing
$ \X \longrightarrow \ti\X, \;
\ups \longrightarrow \ti\ups \;$ and c.c. in (\ref{Sv}).
Now using the result

\bq
K(\F,\ti\X,\F^{*},\ti\X^{*})& =&\;\; K(\F,\X,\F^{*},\X^{*})
+2 Re \;\biggl \{\frac{\exp{(L^{'})} -1}{L^{'}}\;
V^{a}\biggl( \q _{a}(\X) +
Y^{*}_{a}(\F^*,\X)\biggr)\biggr \} \;\; ,\nonumber \\
\ti \ups &=& \ups +\frac{\exp{(L^{'})} -1}{L^{'}}\;V^{a}\q_{a}(\X)
 \;\; ,\\
 L^{'}&\equiv & L_{V \cdot \t^{'}} \;\; , \nonumber
\eq
we are left with the form for the action that couples the
$\s$-model to Yang-Mills fields through the gauging of the isometries:
\be
S\;\;=\;\;
\frac{1}{8}\int {d^4xd^2{\q}d^2\qwt}\; \Biggl (H(\F,\X^{*}) \;+\;
H^{*}(\F^{*},\X) \; + \;
2 Re\; \biggl \{\frac{\exp{(L)} -1}{L}\;V^{a}\;Y^{*}_{a}(\F^*,\X)\biggr \}
\Biggr) \;\; .
\label{Sv1}
\ee
We can still implement a simpler expression for this action if we choose
a Wess-Zumino gauge for the gauge transformations of the component fields
of the vector superfield arising from (\ref{V}) (see for instance
\cite{wessgris}). We also make use of eqs.(\ref{killing1}) and (\ref{kx}).
 In this way, the action (\ref{Sv1}) is rewritten in the following very
simple final form
%\begin{eqnarray}
%S\;\;&=&\;\;
%\frac{1}{8}\int {d^4xd^2{\q}d^2\qwt}\; \Biggl( H(\F,\X^{*}) \;+\;
%H^{*}(\F^{*},\X) \; + \;
%2\; V^{a}\;Y^{*}_{a} \; + \; 2 \;V^{a}\;Y_{a}   \nonumber \\
% & & \;\;\;\;\;\;\;\;\;\;\;
%+ \; 2 \;V^{a}\; V^{b}\; \t_{b}^{i} \; \frac{\pa}{\pa \X^i}
%\;Y^{*}_{a} \;+\;
%2\; V^{a}\; V^{b}\; \t_{b}^{*i} \; \frac{\pa}{\pa \X^{*i}}\;Y_{a}
%\Biggr) \;\; .
%\label{Sv2}
%\end{eqnarray}
\begin{eqnarray}
S\;\;&=&\;\;
\frac{1}{8}\int {d^4xd^2{\q}d^2\qwt}\; \Biggl( H(\F,\X^{*}) \;+\;
H^{*}(\F^{*},\X) \; + \;
2\; V^{a}\;Y^{*}_{a} \; + \; 2 \;V^{a}\;Y_{a}   \nonumber \\
 &&\;\;\; - \; 2 \;V^{a}\; V^{b}\; {\cal K}^I_a \; g_{I \ov J}
 \; {\cal K}^{\ov J}_b
 \Biggr) \;\; .
\label{Sv2}
\end{eqnarray}
Then, we see how the potential $Y_a$, determined in eq.(\ref{y}) for
semi-simple isometry groups, couples to the vector superfield $V^a$
in the gauged action.
As we discussed in the end of Section 3, Abelian factors in the isometry
group may lead to the appearance of arbitrary constants in the potential
$Y_a$. These will also couple to the vector superfield generating the
so called Fayet-Iliopoulos terms \cite{roczek,witten}. In the
general
case of non-semi-simple isometry groups, as it happens in $D=(3+1)$
 dimensions, the potential $Y_a$ may
not be determined, and this will represent an obstruction to the gauging
of the non-linear $\s$-model.

It would be perhaps interesting to consider the possibility of working
with superfield parameters, $\Ld$ and $\G$, that are not real. This
would lead
to the introduction of a family of complex vector superfields to perform
the gauging; however, the appearance of more than one Yang-Mills
multiplet
in the gauging of the isometry group is beyond the scope of the present
work.

\section{Concluding Remarks}
We have here considered a few geometrical aspects concerning non-linear
$\s$-models in the context of an $N=1$ supersymmetry defined in
 $D=(2+2)$. We have shown that such models in general do not need
to be of a K\"ahler
type, even if they are generated by a potential $K$. As an interesting
example, the construction of a real supersymmetric $\s$-model has been
 worked out. Then, restricting ourselves to a special sub-class of
 K\"ahler manifolds, we proceeded with the investigation about the main
points involved in the process of gauging its isometries. In particular,
we have choosen the gauge parameters as constant real superfields, which
would not be possible in a $D=(3+1)$ space-time. We end up with a
superspace action, eq.(\ref{Sv1}), that is invariant under local
isometry transformations. The kinetic terms of $D=(2+2)$ $\s$-models
are off-diagonal (\ref{action1}) and this would signal the presence of
ghosts (negative-norm states) in a space-time of the Minkowski type.
However, the next step would be to carry out a dimensional reduction from
$D=(2+2)$ to $D=(1+2)$ and $D=(1+1)$, where the propagation of fields is
 better controlled. Following the results of \cite{nish} and \cite{oswaldo},
one could go to lower dimensions in such a way that non-physical modes
be eliminated and $\s$-models coupled to Yang-Mills fields may be of some
relevance in connection with conformal theories and integrable models.

The relation of $N=1$ models after dimensional reduction to
 chiral $\s$-models in 2 dimensions \cite{carlos},
and also the construction of an $N=2\;\s$-model in Atiyah-Ward space-time
 will be the subject of further investigation \cite{nosso}.

%The comparison among all possible $N=1$ models after dimensional
%reduction,
%their relation with chiral $\s$-models in 2 dimensions \cite{carlos},
%and also the construction of a $N=2$ model in Atiyah-Ward space-time
 %will be the subject of future work \cite{nosso}.

%\appendix
\section{Appendix}
The K\"ahler space treated in this work is of the type
$\cc^{2m} \times \cc^{\ov {2m}}$ with metric (\ref{metric}), where each
of the blocks is a (2n x 2n) matrix whose respective components
 $g_{i{\ov j}},\;g_{{\hat i}\ov{\hat j}},
\; and\;g_{{\ov i}j},\;g_{{\ov{\hat i}}\hat j}\;$ vanish.
 Since the more general K\"ahler space would allow those
components, our K\"ahler space
is a subclass of the more general one.

{}From (\ref{metric}), we obtain for the connections
\bq
\G^{i}_{jk}= g^{i\ov{\hat r}} \pa_{j} g_{k\ov{\hat r}}\; ,
\nonumber \\
\G^{\hat i}_{{\hat j}{\hat k}}= g^{{\hat i}\ov r}\pa_{\hat j}
g_{{\hat k}\ov r}\;, \nonumber \\
\G^{\ov i}_{{\ov j}\ov k} = g^{{\ov i}{\hat r}}\pa_{\ov j}
g_{{\ov k}{\hat r}}\;,
\nonumber \\
\G^{\ov {\hat i}}_{{\ov {\hat j}}{\ov {\hat k}}} =
g^{{\ov {\hat i}}r}
\pa_{\ov {\hat j}}g_{{\ov {\hat k}}r}\; ,
\eq
and for the curvatures
\bq
\left.
\begin{array}{l}
{\cal R}^{i}_{jkL} = \pa_{L}\G^{i}_{jk}\;\;\; \mbox{with}\;\;\
L=\{{\hat l},{\ov l},{\ov {\hat l}}\}\; ,\;\;\;\;\;\;
{\cal R}^{i}_{jKl} = -\pa_{K}\G^{i}_{jl}\;\;\; \mbox{with}
\;\;\; K=\{{\hat k},{\ov k},{\ov {\hat k}}\}\; , \nonumber \\
{\cal R}^{\hat i}_{{\hat j}{\hat k}L} =
\pa_{L}\G^{\hat i}_{{\hat j}{\hat k}}
\;\;\; \mbox{with}\;\;\; L=\{l,{\ov l},{\ov {\hat l}}\}\;,\;\;\;\;\;\;
{\cal R}^{\hat i}_{{\hat j}K{\hat l}} =
 -\pa_{K}\G^{\hat i}_{{\hat j}{\hat l}}
\;\;\; \mbox{with}\;\;\; K=\{ k,{\ov k},{\ov {\hat k}}\}\;,
 \nonumber \\
{\cal R}^{\ov i}_{{\ov j}{\ov k}L} = \pa_{L}\G^{\ov i}_{{\ov j}\ov k}\;\;\;
\mbox{with}\;\;\; L=\{l,{\hat l},{\ov{\hat l}}\}\;,\;\;\;\;\;\;
{\cal R}^{\ov i}_{{\ov j}K{\ov l}}=-\pa_{K}\G^{\ov i}_{{\ov j}\ov l}\;\;\;
\mbox{with}\;\;\; K=\{k,{\hat k},{\ov {\hat k}}\}\;, \nonumber \\
{\cal R}^{\ov {\hat i}}_{{\ov {\hat j}}{\ov {\hat k}}L} =
\pa_{L}\G^{\ov {\hat i}}_{{\ov {\hat j}}{\ov {\hat k}}}\;\;\;
\mbox{with}\;\;\;L=\{l,{\ov l},{\hat l}\}\;,\;\;\;\;\;\;
{\cal R}^{\ov {\hat i}}_{{\ov {\hat j}}K{\ov {\hat l}}} =
-\pa_{K}\G^{\ov {\hat i}}_{{\ov {\hat j}}{\ov {\hat l}}}\;\;\;
\mbox{with}\;\;\;K=\{k,{\ov k},{\hat k}\}\;.
\end{array} \right.
\eq

Now, we shall analyse the assumption on the structure of the
 Killing vectors shown in eq.(\ref{killing1}).
We intend to show just a sketch of a proof that is
in complete analogy to the one given in \cite{yano2},
 so that it will be just a
slight modification of the Theorems 2.4 and 2.5 of that reference.

As it is well known, in a compact K\"ahler space a necessary and suficient
condition for a contravariant vector ${\cal K}^{I}$ be a Killing vector is
\bq
g^{JK}\nabla_{J}\nabla_{K}{\cal K}^{I} + {\cal R}^{I}_{J}{\cal K}^{J}=0
\;\; ,
\nonumber \\
\nabla_{I}{\cal K}^{I}=0 \;\; ,
\label{killing2}
\eq
where ${\cal R}^{I}_{J}$ is the Ricci tensor.

Let us impose that the Killing vector ${\cal K}^{I}=(k^{i},k^{\hat i},
k^{\ov i},
k^{\ov{\hat i}})$
satisfies
\bq
 \nabla_{i} k^{i}=\nabla_{\hat i} k^{\hat i}=0 \;\;and \;\;c.c. \;\; .
\label{derivada}
\eq
Then,
from (\ref{killing2}), we also have
$\z^{I} = (k^i,0,0,0),\;\t^{I} = (0,k^{\hat i},0,0),\;\l^{I}=(0,0,k^{\ov i},0)$
and $\h^{I}=(0,0,0,k^{\ov{\hat i}})$ as Killing vectors.
This allows us to write for each of them,
\bq
\nabla_{I}\z_{J}+\nabla_{J}\z_{I}=0 \;\; etc. \;\; ,
\label{killing3}
\eq
with $\z_{I}=(0,0,0,k_{\ov{\hat i}}),\;\;k_{\ov{\hat i}}=
g_{{\ov{\hat i}}j}k^{j}$. Recalling that
$\G^{\ov {\hat k}}_{{\ov {\hat i}}{\ov {\hat j}}}$
is the only non-vanishing component of $\G^{\ov {\hat k}}_{IJ}$, we have
from (\ref{killing3}) that
$\z_{\ov{\hat i}}= \z_{\ov{\hat i}}(\X^*)$ or $k_{\ov{\hat i}}=
k_{\ov{\hat i}} (\X^*)$, and in an analogous way $k_{i}=k_{i}(\F),
\; k_{\hat i}=k_{\hat i}(\X)$ and
$k_{\ov i}=k_{\ov i}(\F^*)$. Those covariant components of the Killing
vector ${\cal K}^{I}$ being holomorphic, we have from \cite{yano2} that
${\cal K}^{I}$ is harmonic, i.e., it satisfies,
\bq
\nabla_{I}{\cal K}_{J}-\nabla_{J}{\cal K}_{I}=0 \;\;.
\label{harmonico}
\eq
Since ${\cal K}^{I}$ is a Killing vector we also have
$\nabla_{I}{\cal K}_{J}+\nabla_{J}{\cal K}_{I}=0$. This, together with
eq.(\ref{harmonico}), gives
$\nabla_{I}{\cal K}_{J}=0$,
and then $\nabla_{I}{\cal K}^{J}=0$, which also implies
\bq
k^{i}=k^{i}(\F),\;k^{\hat i}=k^{\hat i}(\X),\;k^{\ov i}=k^{\ov i}(\F^*),
\;k^{\ov{\hat i}}=k^{\ov{\hat i}}(\X^*)\;\;.
\label{holomorphico}\eq
 We have  then proven  that
 Killing vectors satisfying (\ref{derivada}) are holomorphic in all their
coordinates.

Conversely, let ${\cal K}^{I}$ be a vector satysfying (\ref{derivada})
and holomorphic in all its coordinates (\ref{holomorphico}). From the Ricci
 identities
\bq
\nabla_{J}\nabla_{K} {\cal K}^{I} - \nabla_{K}\nabla_{J} {\cal K}^{I}
={\cal R}^{I}_{LKJ}{\cal K}^{L} \;\; ,
\eq
we get
\bq
\nabla_{\ov{\hat j}}\nabla_{k} k^{i}= {\cal R}^{i}_{lk{\ov{\hat j}}}\,k^{l}
\;\; , \nonumber \\
\nabla_{\ov j}\nabla_{\hat k} k^{\hat i}= {\cal R}^{\hat i}_{{\hat l}
{\hat k}{\ov j}}\,k^{\hat l}  \;\; , \nonumber \\
\nabla_{\hat j}\nabla_{\ov k} k^{\ov i} = {\cal R}^{\ov i}_{{\ov l}
{\ov k}{\hat j}}\,k^{\ov l} \;\; , \nonumber \\
\nabla_{j}\nabla_{\hat {\ov k}} k^{\ov {\hat i}} =
{\cal R}^{\ov {\hat i}}_{{\ov {\hat l}}
{\ov {\hat k}}j}\,k^{\ov {\hat l}} \;\; .
\eq
Contracting each of them respectively with $g^{{\ov {\hat j}}k},\;g^{{\ov j}
{\hat k}},\;g^{{\hat j}{\ov k}},\;g^{j{\ov {\hat k}}}$, and using
(\ref{derivada}), we can write
\bq
&& g^{{\ov {\hat j}}k}\nabla_{\ov {\hat j}}\nabla_{k} k^{i} +
{\cal R}^{i}_{l}\,k^{l} = 0 \;\;\; , \;\;\; \nabla_{i} k^{i}=0
\;\;\; and \;\;\;  c.c. \;\; , \nonumber \\
&& g^{{\ov j}{\hat k}}\nabla_{\ov j}\nabla_{\hat k} k^{\hat i} +
{\cal R}^{\hat i}_{\hat l}\,k^{\hat l} = 0\;\;\; , \;\;\;
\nabla_{\hat i} k^{\hat i}=0\;\;\; and \;\;\;
c.c.\;\; ,
\eq
or in a compact way,
\bq
g^{JK} \nabla_{J}\nabla_{K} {\cal K}^{I} + {\cal R}^{I}_{L} {\cal K}^{L}
=0 \;\;\;\mbox{and} \;\;\;\nabla_{I}{\cal K}^{I}=0\;\;. \nonumber
\eq
This is exactly the condition (\ref{killing2}) for a Killing vector.
We have proven then that a vector satisfying (\ref{derivada}) is
a Killing vector if and only if its components are holomorphic in
all coordinates $\F, \;\X , \;\F^*,\;\X^*$.

\section*{Acknowledgements}

\small
The authors are deeply indebted to Oswaldo M. Del Cima and
Marco A. de Andrade for several explanations on technical aspects
concerning Susy
in $D=(2+2)$. They would also like to thank Dr. L.P. Colatto for
 drawing
their attention to some useful references. $C.N.P_{q}$ is
acknowledged
for invaluable financial help.

\end{document}